\documentstyle[aps,epsfig,manuscript,multicol]{revtex}
\begin{document}
\newcommand{\beq}{\begin{equation}}
\newcommand{\eeq}{\end{equation}}
\newcommand{\sg}{\sigma_p}
\newcommand{\sm}{\sigma_c}
\newcommand{\lf}{\langle L_f \rangle}
\newcommand{\la}{\langle L_a \rangle}
\newcommand{\tf}{{\rm Tr}_f}
\newcommand{\nt}{N_\tau}
\newcommand{\ns}{N_\sigma}
\newcommand{\f}{\beta_f}
\newcommand{\ba}{\beta_a}
\newcommand{\bv}{\beta_v}
\newcommand{\bt}{\beta}
\newcommand{\bvc}{\beta_{vc}}
\newcommand{\lm}{\lambda}
\title{
Establishment of a stable vortex state in the peak effect region
in a weakly pinned  superconductor CeRu$_2$}

\author{A.A. Tulapurkar$^1$\footnote{E-mail:ashwin@tifr.res.in},D. Heidarian$^1$,
S. Sarkar$^1$, S. Ramakrishnan$^1$, A.K. Grover$^1$, E.
Yamamoto$^2$, Y. Haga$^2$, M. Hedo$^3$, Y. Inada$^3$ and Y.
Onuki$^{2,3}$}
\address{$^1$ Department of Condensed Matter Physics and Materials Science,
Tata Institute of Fundamental Research, Colaba, Mumbai 400005,
India}
\address{$^2$ A.S.R.C. Japan Atomic Energy Research
Institute, Tokai, Ibaraki 319-11, Japan}
\address{$^3$ Faculty of Science, Osaka University, Toyonaka 560,
Japan}
\maketitle
\begin{abstract}

We present magnetization data on a weakly pinned CeRu$_2$ single
crystal showing the existence of a stable state of the vortex
lattice in the peak effect region. The stable state is achieved by
cycling the magnetic field by small amplitude. This stable state
is characterized by a unique value of critical current density
($J_c$), independent of the magnetic history of the sample. The
results support the recent model proposed by  Ravikumar {\it et
al} \cite{r39} on the history dependence of the $J_c$.
\end{abstract}
\vskip5mm
\section{Introduction}
The critical current density $J_c$  of a type II superconductor
usually decreases monotonically with increasing $H$ or $T$.
However, in weakly pinned superconductors, the interplay between
the intervortex interaction and the flux pinning produces an
anomalous peak in $J_c$ before it decreases to zero at the
superconductor-normal phase boundary ($H_{c2}/T_c(H)$ line)
\cite{r11}. This effect is known as the Peak Effect (PE) and it
signifies that the vortex phase undergoes a transition from an
ordered phase to a disordered phase \cite{r11,r16,r17}.

The PE in weakly pinned superconductors is accompanied by
metastable vortex states. Each metastable vortex configuration is
characterized by a different $J_c$, which depends on the
thermomagnetic history of the superconductor.  Bean's critical
state model (CSM), which is usually used to describe the
hysteritic magnetic response of the superconductors assumes a
unique value of $J_c$, independent of the thermomagnetic history
\cite{r38}. Therefore, it fails to account for the magnetic
properties of superconductors in the PE region. Recently,
Ravikumar et al proposed a model to explain the history dependence
of the critical current density \cite{r39}. This model postulates
a stable state of vortex lattice with a critical current density
$J_c^{st}$ determined uniquely by the field and temperature,
independent of the past magnetic history. This stable state can be
reached from any metastable vortex state by cycling the applied
field by small amplitude.

We report here the results of an investigation of the stable state in the
PE region in a single crystal sample of CeRu$_2$.

\section{Phenomenological model for history effects and metastablity in weakly pinned
superconductors} Ravikumar {\it et al} have proposed a model for
the history dependence of the critical current density \cite{r39}.
An important assumption of this model is the existence of a stable
vortex state with a critical current density $J_c^{st}$ which is
unique for a given temperature and field. They postulated the
following equation to describe how the vortex state evolves from
one configuration to other:
\begin{equation}
J_c(B+\Delta B)=J_c(B) + (|\Delta B|/B_r)(J_c^{st}-J_c)
\end{equation}
$B_r$ is a macroscopic measure of the metastability and describes
how strongly $J_c$ is history dependent. Physically, we may
imagine that in the absence of thermal fluctuations, it is the
change in local field $B$ that can move the vortices from their
metastable configuration to a nearby stable state. It can be seen
from the above equation that a metastable vortex state with $J_c
\neq J_c^{st}$ can be driven into the stable state by cycling the
field by small amplitude.

\section{Experimental Results and discussion}
DC magnetization measurements have been carried out using a
12~Tesla Vibrating Sample Magnetometer (VSM) [Oxford Instruments,
U.K.] on a single crystal sample of CeRu$_2$ ($T_c\sim$ 6.2~K).
The sample was vibrated with an amplitude of 1.5 mm and a
frequency of 55 Hz. The dc field was swept at rate of 0.02
Tesla/min., while recording the magnetization hysteresis loops at
4.5~K. All the measurements were carried out by cooling the sample
to 4.5~K and then applying the magnetic field (ZFC mode). Fig.~1
shows the hysteresis loop, constituting $M vs. H$ curve in the
increasing (forward) and decreasing (reverse) field cycles
(envelope curve). According to Bean's critical state model, this
hysteresis in magnetization $\Delta M(H) =
M(H\uparrow)-M(H\downarrow)$ provides a measure of the critical
current density ($J_c \propto \Delta M$) \cite{r29}. Thus, an
anomalous increase in the width of the hysteresis loop is a
distinct indicator of the occurrence of the PE. We identify
$H_{pl}^+$ as the onset field of the PE on the forward leg, where
$M$ begins to decrease sharply. The onset field for the PE on the
reverse leg is marked as $H_{pl}^-$ which is less than $H_{pl}^+$.
The field where the magnetization hysteresis bubble is the widest
identifies the peak field $H_p$ and the collapse of the hysteresis
loop locates the irreversibility field $H_{irr}$, above which the
critical current density falls below the measurable limit of the
method used.

Fig.~\ref{MHL} also shows the minor hysteresis loops (MHL)
starting from both forward and reverse legs recorded at different
fields within the PE region. As per CSM, all the MHLs are expected
to lie within the envelope curve and saturate by reaching the
envelope curve. However, it can be seen from the Fig.~\ref{MHL}
that the MHL starting from a point $H_{pl}^+ <H < H_p$ on the
forward cycle saturate without meeting the reverse curve although
they remain well within the envelope curve. On the other hand,
MHLs starting from a point $H_{pl}^- <H < H_p$ on the reverse
curve overshoot the forward curve. These observations can not be
understood within the CSM and indicate that $J_c$ is magnetic
history dependent over some part of the PE region. Moreover, they
also imply that the critical current density in the forward cycle
($J_c^{for}$) is less than the critical current density in the
reverse cycle ($J_c^{rev}$). Using the saturation points of the
MHLs, magnetization hysteresis loops corresponding to $J_c^{for}$
and $J_c^{rev}$ can be generated. $J_c^{for}$ and $J_c^{rev}$ can
be obtained from the width of the hysteresis loops. (See
Fig.~\ref{Jc})

In Fig.~\ref{cycling}, we show the MHLs obtained by repeatedly
cycling the field starting from 2 points ($H_{pl}^- <H < H_p$) on
the reverse cycle. The cycling amplitude $\Delta H$ is chosen such
that it is above the threshold field required to reverse the
direction of shielding currents throughout the sample. From the
Fig.~\ref{cycling}, it is clear that the MHLs show shrinkage with
field cycling and after few field cycles, the MHLs retrace each
other indicating that $J_c$ does not change further with field
cycling. We, therefore, conclude that the vortex state is in a
stable configuration. On the forward cycle, the MHLs ($H_{pl}^+ <
H < H_p$) show expansion with field cycling and finally the MHLs
retrace each other indicating the stable vortex state.
Fig~.\ref{magn} shows the stable magnetization hysteresis loop
obtained from the stable MHLs. The critical current density in the
stable state ($J_c^{st}$) can be obtained from the width of the
stable state hysteresis loop. Fig.~\ref{Jc} shows the critical
current density in the forward and reverse cycles and in the
stable state. It can be seen that they follow the relation:
$J_c^{rev} > J_c^{st}
> J_c^{for}$.
\section{Conclusions}

In this paper, we have studied different metastable vortex
configurations in the PE region of the weakly pinned CeRu$_2$.
These states have been  characterized by different values of
critical current densities, which were obtained by magnetization
measurements. As an important part, we have shown the existence of
a stable state of the vortex array in the PE region, which is
achieved by cycling the field by small amplitudes.

\begin{figure}
\caption{Magnetization hysteresis loop in the PE region at 4.5~K.
Panel (a) shows the minor hysteresis loops (MHLs) starting from
the forward leg and panel (b) shows MHLs starting from the reverse
leg. }
 \label{MHL}
\end{figure}

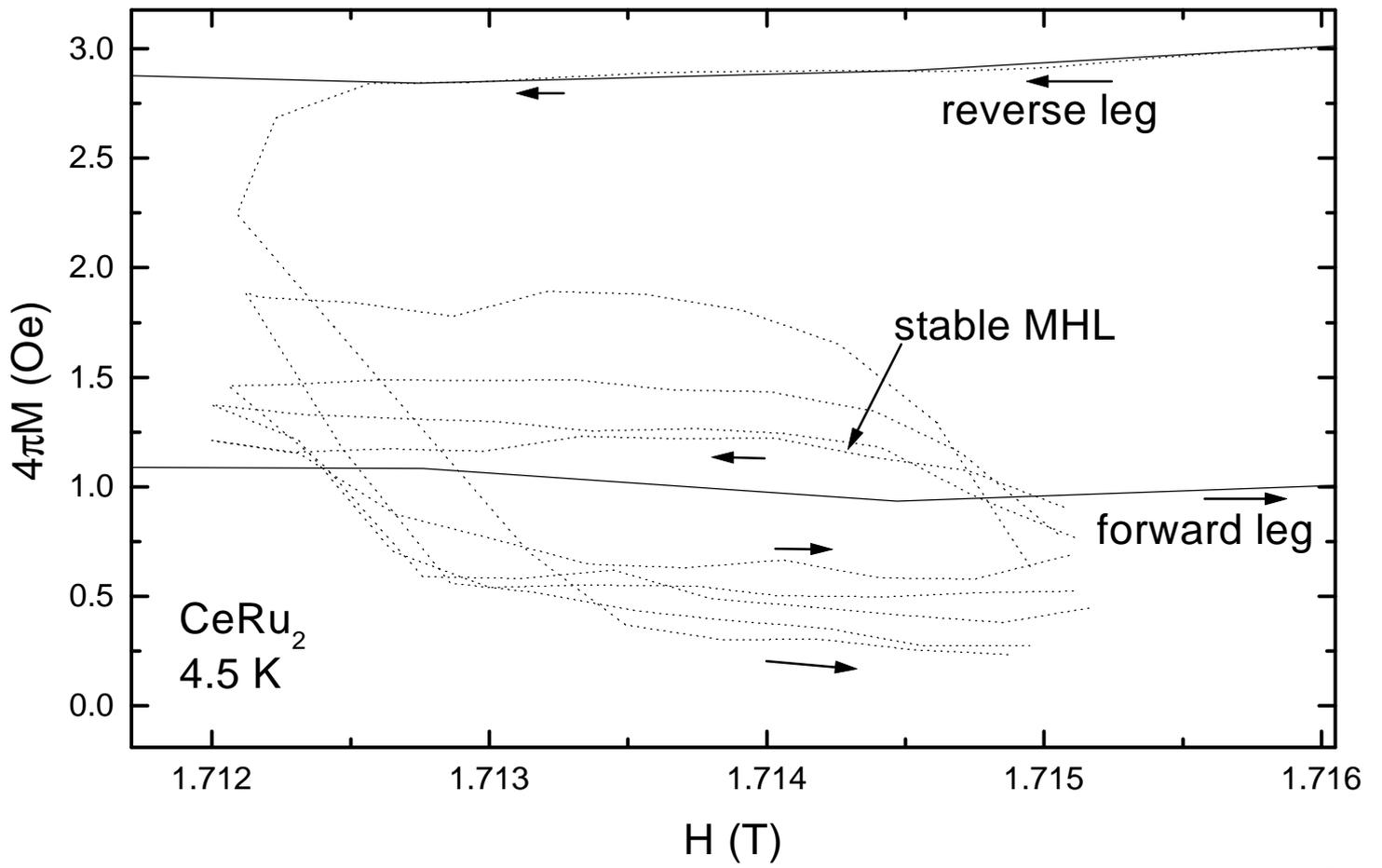
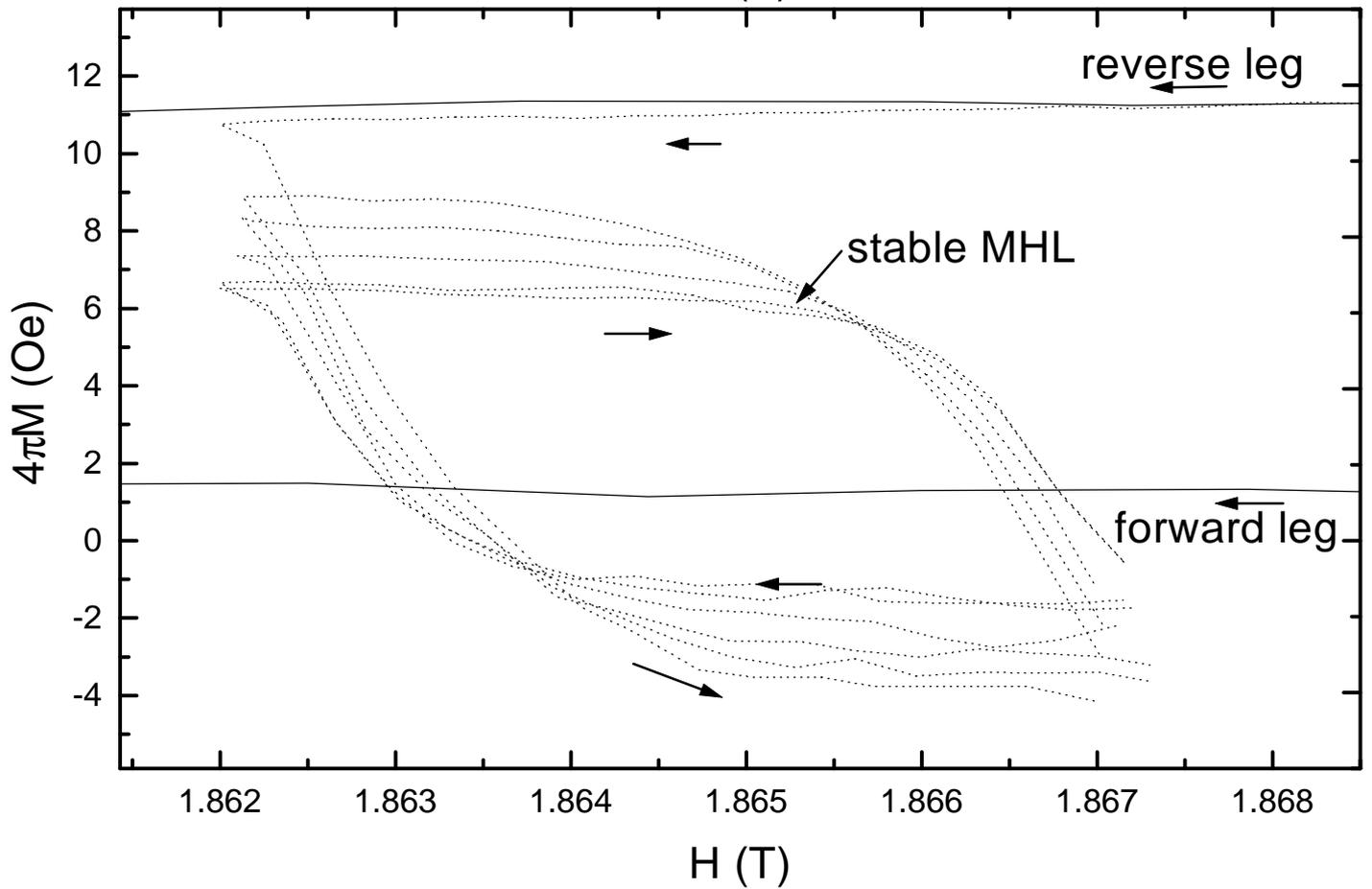
\begin{figure}
\caption{Minor hysteresis loops obtained by cycling the field in
the PE region starting from 2 points on the reverse leg. The inner
most MHL corresponds to the stable state.} \label{cycling}
\end{figure}

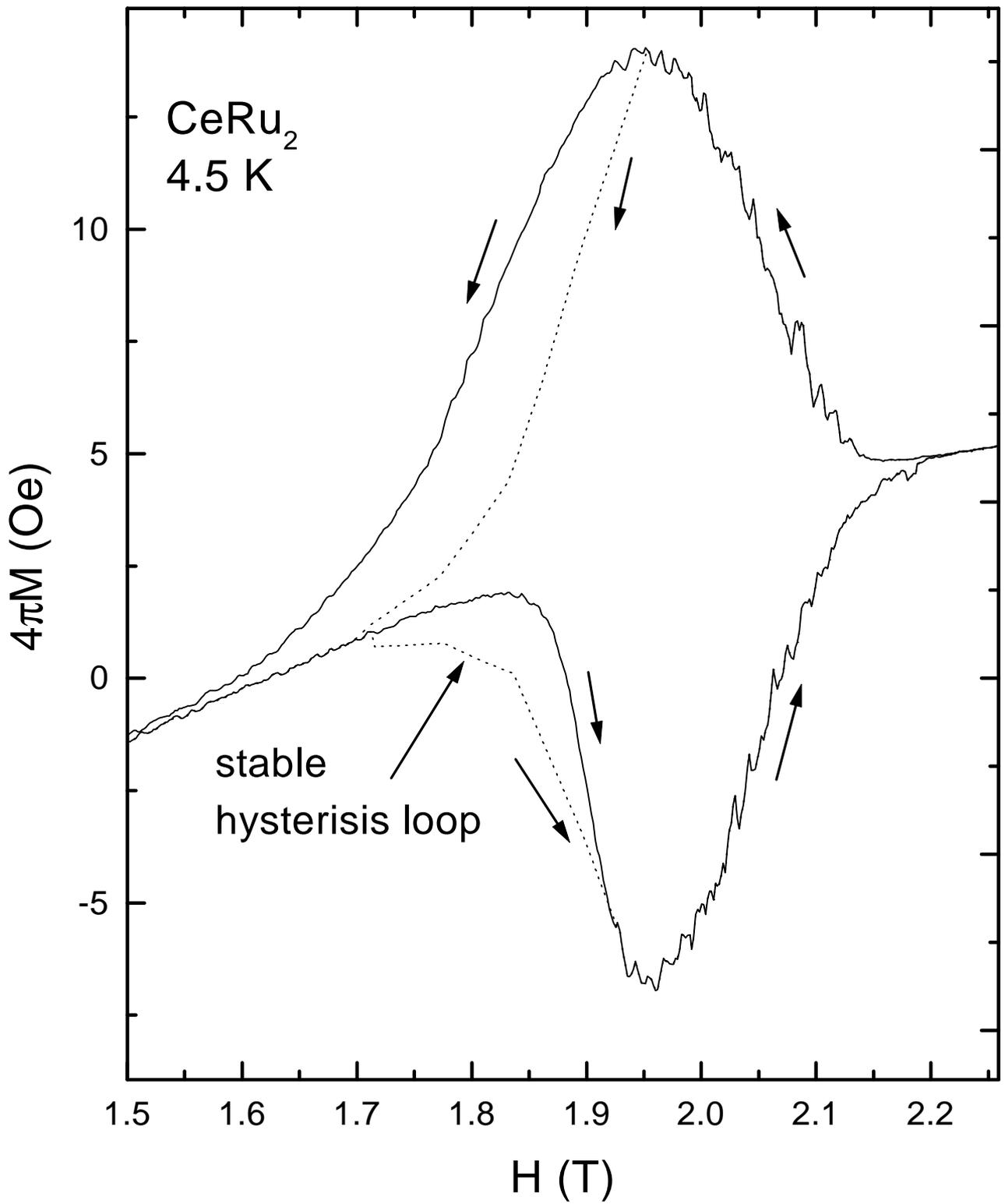
\begin{figure}
\caption{Magnetization hysteresis loop of CeRu$_2$ in the PE
region at 4.5~K (continuous line). Dotted line is the saturated
magnetization curve obtained after repeated field cycling
(magnetization of the stable state)} \label{magn}
\end{figure}

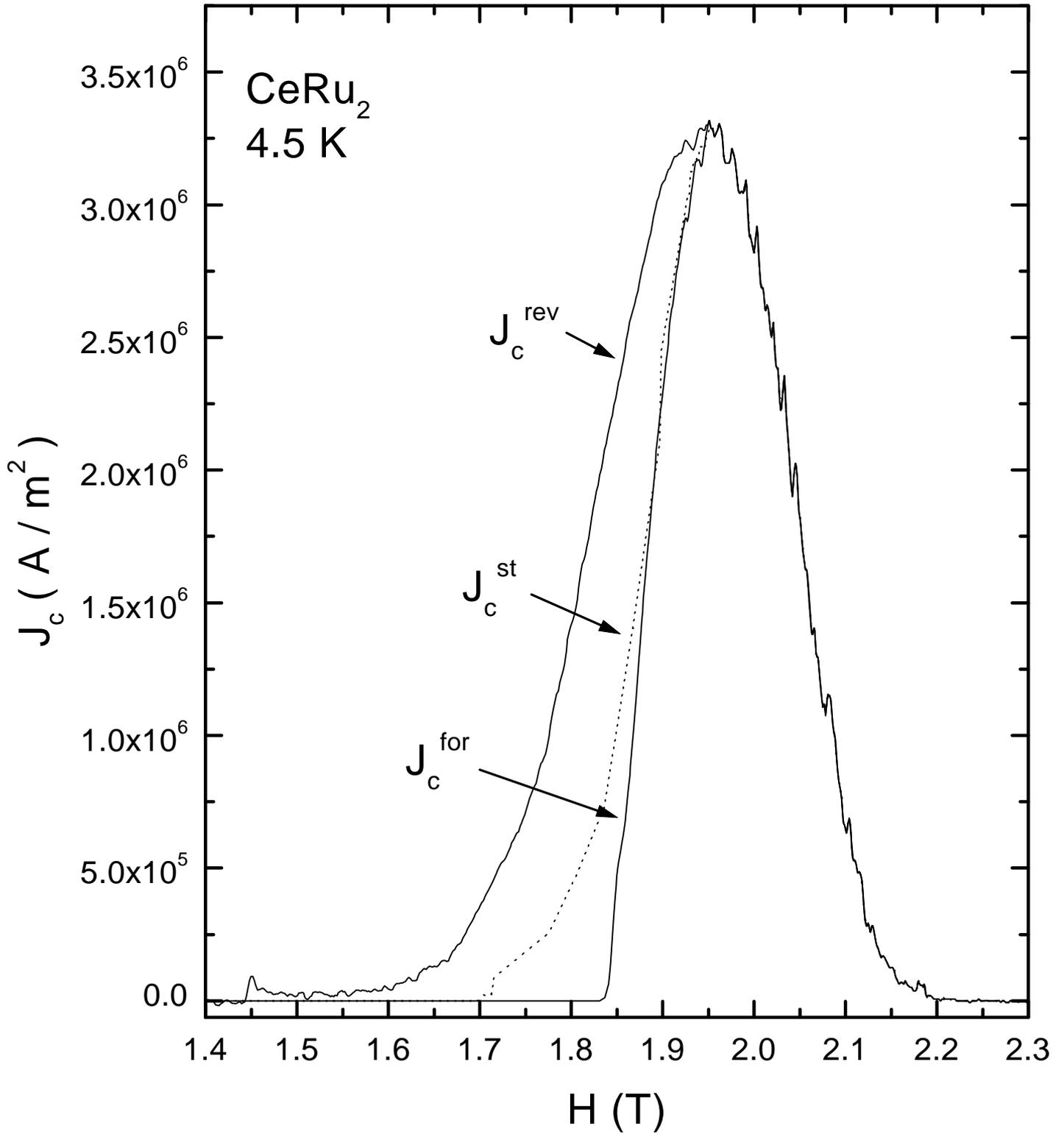
\begin{figure}
\caption{Critical current densities $J_c^{for}$ and $J_c^{rev}$ in
the increasing and decreasing field cases respectively (continuous
line). Dotted line shows the stable state critical current
density.} \label{Jc}
\end{figure}

\end{document}